\documentclass[preprint,showpacs,preprintnumbers,amsmath,amssymb]{revtex4}
\usepackage{graphicx}
\usepackage{dcolumn}
\usepackage{bm}
\usepackage{psfig}  
\usepackage{epsfig}

\begin{document}

\newcommand{\vk}{{\vec k}}
\newcommand{\vK}{{\vec K}} 
\newcommand{\vb}{{\vec b}} 
\newcommand{{\vp}}{{\vec p}} 
\newcommand{{\vq}}{{\vec q}} 
\newcommand{\vx}{{\vec x}}
\newcommand{\tr}{{{\rm Tr}}} 
\newcommand{\beq}{\begin{equation}}
\newcommand{\eeq}[1]{\label{#1} \end{equation}} 
\newcommand{\half}{{\textstyle \frac{1}{2} }}
\newcommand{\lton}{\mathrel{\lower.9ex \hbox{$\stackrel{\displaystyle
<}{\sim}$}}} 
\newcommand{\gton}{\mathrel{\lower.9ex \hbox{$\stackrel{\displaystyle
>}{\sim}$}}} 
\newcommand{\ee}{\end{equation}}
\newcommand{\bea}{\begin{eqnarray}} 
\newcommand{\eea}{\end{eqnarray}}
\newcommand{\beqar}{\begin{eqnarray}} 
\newcommand{\eeqar}[1]{\label{#1}\end{eqnarray}} 
\newcommand{\pleft}{\stackrel{\leftarrow}{\partial}}
\newcommand{\pright}{\stackrel{\rightarrow}{\partial}}

\begin{flushright}
\vskip .5cm
\end{flushright} \vspace{1cm}

\title{TRANSVERSE MOMENTUM DIFFUSION AND BROADENING OF THE 
BACK-TO-BACK \\ DI-HADRON CORRELATION FUNCTION}

\author{Jianwei Qiu}%
 \email{jwq@iastate.edu}

\author{Ivan Vitev}
 \email{ivitev@iastate.edu}

\affiliation{Iowa State University, Department of Physics and Astronomy 
\\  \hspace*{3cm}  Ames, IA 50011, USA\hspace*{3cm}  }%

\begin{abstract}
We extend the Gyulassy-Levai-Vitev reaction operator approach to multiple 
elastic scattering of fast partons traversing dense nuclear matter to
take into account the leading power corrections due to the medium recoil
and to derive the change in the partons' longitudinal momentum.
We employ a boost-invariant formalism to generalize previous treatments 
of the problem, which were specific to the target rest frame. We apply 
the transverse momentum diffusion results in a simple analytic model to 
evaluate the broadening of the back-to-back di-hadron 
correlation function in $d+Au$ reactions.
\end{abstract}



\maketitle

\section{Introduction}

New  experiments at the  Relativistic Heavy Ion Collider (RHIC) 
will provide high statistics measurements of photons, leptons and hadrons 
at large transverse momentum ($p_T$) in high energy nucleus-nucleus 
collisions~\cite{Adler:2003qi}.  Particle production from 
a single hard scattering with momentum exchange much larger than 1/fm 
should be localized in space-time. It is multiple parton scattering before 
or after the hard collision that is sensitive to the properties of the 
nuclear matter~\cite{Luo:ui,Baier:1997kr,Gyulassy:2000fs}. By comparing 
the high-$p_T$ observables~\cite{Vitev:2002pf,Arleo:2002kh,Gyulassy:2000gk}   
in $p+p$, $d+A$ and $A+A$ reactions, we are able to study the strong 
interaction dynamics of QCD in the vacuum, cold nuclear matter and hot
dense medium of quarks and gluons, respectively.

In this Letter, we extend the Gyulassy-Levai-Vitev (GLV) reaction 
operator approach to multiple elastic scatterings~\cite{Gyulassy:2002yv} 
to take into account the longitudinal momentum reduction for a hard jet 
($\sqrt{p^2}/p^0 \simeq 0$) that  propagates through dense nuclear matter
due to the medium recoil. At the leading power approximation in 
${\cal O}(1/P)$ for a fast parton of momentum $p^\mu = (P,{\bf 0}_\perp,P)$ 
elastic scattering in nuclear matter introduces transverse momentum 
broadening without changing its ``+'' light-cone component.  By 
including leading power corrections due to the medium response, we 
evaluate the parton's longitudinal momentum shift, 
$- \Delta p_\parallel   =  \langle {\bf p}_\perp^2\rangle / 2P$, 
that ensures 3D momentum conservation in an elastic collision up to  
corrections of ${\cal O}(\langle  {\bf p}^2_\perp \rangle^2/P^2)$.
In our derivation we introduce a boost invariant formulation of the 
problem of multiple jet interactions that generalizes the Gyulassy-Wang 
picture of static or massive scattering centers~\cite{Gyulassy:1993hr} 
employed in recent studies~\cite{Baier:1997kr,Gyulassy:2000fs,Gyulassy:2002yv}.

We apply the transverse momentum diffusion results in a simple 
analytic model to evaluate the nuclear modification of the back-to-back 
di-hadron  correlation function  $C(\Delta \phi) = 
N^{h_1,h_2}(\Delta \phi)/N_{tot}^{h_1,h_2}$ 
in $d+Au$ reactions at RHIC using cold nuclear matter transport 
coefficients~\cite{Vitev:2002pf}  extracted from low energy $p+A$ 
data~\cite{Cronin:zm}. We find a small, weakly dependent on  centrality, 
increase of the far-side ($\Delta \phi > \pi/2$)  width $\sigma_{Far}$ 
of  $C(\Delta \phi)_{dAu}$.  Such behavior is distinctly different 
from the reported disappearance of the back-to-back 
correlations in central $Au+Au$ reactions~\cite{Adler:2002tq}, which can 
be understood  in terms of strong final state radiative energy 
loss~\cite{Hirano:2003hq} and subsequent redistribution of the lost energy 
in the parton system~\cite{Pal:2003zf}.

\section{Multiple Elastic Scattering In The GLV Reaction Operator Approach}

The GLV reaction operator formalism was developed for calculating 
the induced radiative energy loss of  hard quarks or gluons when they 
pass through a dense medium~\cite{Gyulassy:2000fs}. In this approach 
the multiparton dynamics is described by a series expansion in 
$\chi = \int \sigma_{el}(z) \rho(z) dz = L/\lambda$, the mean number 
of interactions that a fast projectile undergoes along its trajectory. 
Each interaction is represented by a reaction operator that summarizes 
the unitarized basic scattering between the propagating system and  
the medium. The summation to all orders in $\chi $ is achieved 
by a recursion of the reaction operators and is given in a closed form 
in~\cite{Gyulassy:2000fs,Gyulassy:2002yv}.

In this Letter we focus on the case of multiple elastic (no-radiation) 
scattering of a jet of momentum $p$ in nuclear matter.
Let $M_0 = i  e^{ip \cdot z_0} j(p) \, {\bf 1}_{d_R \times d_R}$  be 
the amplitude of the parton in color representation $R$ of dimension 
$d_R$ prepared at a position $z_0^\mu = [z^+_0,z^-_0,{\bf z}_{0\;\perp}]$, 
where the light-cone coordinates are defined as 
$z^{\pm}=(z^0\pm z^3)/\sqrt{2}$. The unperturbed inclusive distribution 
of jets in the wave packet is given by~\cite{Gyulassy:2002yv}: 
\beq
d^3 N^i= {\rm Tr}\;|M_0|^2 \frac{d^3 \vec{\bf p} }{(2\pi)^3\, 2p^0}
=  |j(p)|^2 d_R\, \frac{dp^+  d^2 {\bf p}_\perp}{2p^+\,(2\pi)^3 } \;\; . 
\eeq{n0} 
For the instructive case of a normalized forward monochromatic beam 
($p_0=p_\parallel \equiv P$) 
\beq   
\frac{d^3N^i}{d p^+ d^2 {\bf p}_\perp } |_{p^- =  
\frac{{\bf p}_\perp^2}{2 p^+ }}= 
\delta(p^+ - \sqrt{2} P) \delta^2({\bf p}_\perp )  \;\;.
\eeq{fbeam}
In Eq.(\ref{fbeam}) the constraint on the parton's ``$-$'' light-cone 
component  from the $p^2=0$ on-shell condition  is also shown.

In the presence of nuclear matter the multiple jet interactions 
are modeled by scattering in the presence of an external non-Abelian 
field $V^{\mu,c} (q)$ which satisfies the condition: 
$q_\mu V^{\mu,c} (q)=0$, $c$ being the color index. When the parton energy  
is much larger than the typical momentum scale of the medium, 
$P^2 \gg |q^2|$, we have  
\beq
V^{\mu,c} (q) = n^{\mu}\, 2\pi 
\delta(q^+)\, V^c (q)\, e^{i q \cdot z} \;, \qquad 
 g_s V^c(q) \equiv v(q) \, T^c(T) \;\;, 
\eeq{V-solution}
where $g_s$ is the strong coupling constant, the four-vector
$n^\mu = \delta^{\mu,-} = [0,1,{\bf 0}_\perp ]$ and 
$q^\mu = [ q^+=0,q^-,{\bf q}_\perp ]$ is the momentum exchange with
the medium.  The phase in Eq.(\ref{V-solution}) keeps track of the 
position of the scatterer relative to a fixed space-time point, e.g. 
the hard collision vertex. The color matrix $T^c(T)$ in Eq.(\ref{V-solution}) 
represents the non-Abelian charge of the field radiated by a quark 
(or a gluon) with $T$ = fundamental (or adjoint) representation of $SU_c(N)$ 
of dimension $d_T$.  The Fourier transform  of the non-Abelian field is given 
by $v(q)$ and, similarly to the Gyulassy-Wang model~\cite{Gyulassy:1993hr}, 
we employ the color-screened Yukawa type but with Lorentz boost invariance: 
\beq
v(q) \equiv  \frac{4\pi \alpha_s}{-q^{\;2}+\mu^2} 
= \frac{4\pi \alpha_s}{{\bf q}_\perp^{\;2}+\mu^2}  =v({\bf q}_\perp) \;\; ,
\eeq{vq}
where we have used the $q^+=0$ choice of frame. This specific form 
of $v(q)=v({\bf q}_\perp)$ is particularly useful since in-medium  
interactions in both hot and cold nuclear matter are  of finite range 
$r_{int.}=\mu^{-1}$ and we shall assume that $\lambda \mu \gg 1$, where 
$\lambda$ is the parton's mean free path.

\begin{figure}[t]
\begin{center} 
\vspace*{-1.in}
\psfig{file=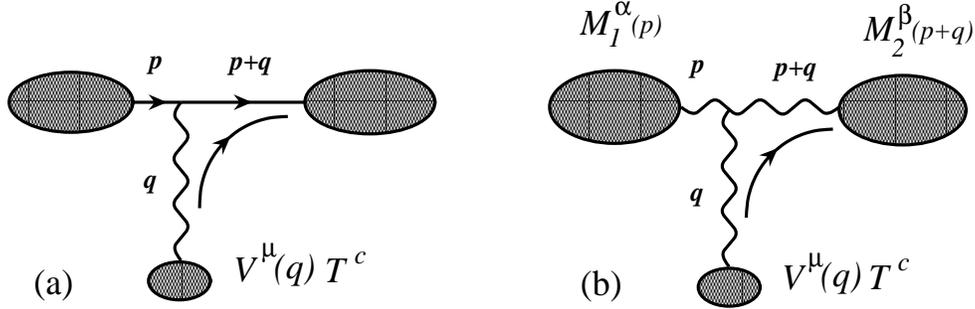,height=4.in,width=5.4in,angle=0}
\vspace*{-1.4in}
\caption{In-medium interaction of fast almost on-shell $p^2 \simeq 0$ 
quarks and gluons with an external color field $V^\mu(q)\, T^c$ located 
at position $z$. The momentum flow for the elastic scattering subprocess 
is shown in the diagrams.}
\label{ampl:fig1}
\end{center} 
\end{figure}

Figure~\ref{ampl:fig1} represents the simplest case of a quark or gluon 
jet with a large forward momentum $p^+$  scattering in the medium. For 
initial state interactions, the elastic scattering amplitude for the subprocess 
with momentum flow illustrated in Fig.~\ref{ampl:fig1}(a) is given by:   
\beqar
M_{el}^q &=& \int \frac{d^4 q}{(2\pi)^4} \; \frac{1}{2} 
\tr \left[ \frac{i \gamma  \cdot ( p + q )}{ (p+q)^2 + i \epsilon } 
( -ig_s \gamma^\mu T^c(F) )  \right]  \, V^c_\mu(q)  
\nonumber \\[1ex]
& = & \int \frac{d q^-}{2\pi}\frac{d^2 {\bf q}_\perp}{(2\pi)^2} 
\left[\frac{2p^+ \, v({\bf q}_\perp)}{(p+q)^2+i\epsilon} \right]
e^{iq^-(z-z_0)^+} e^{-i{\bf q}_\perp \cdot ({\bf z}_\perp-{\bf z}_{0\,\perp})}
 \, T^c(F) \; T^c(T) \;\;,  
\eeqar{q-el-amp} 
where  $T^c(F)$ is the color matrix in the quark representation.
Before giving the corresponding matrix element for the gluon case 
in Fig.~\ref{ampl:fig1}(b) we specify the use of the light-cone gauge 
$A_\mu n^\mu=A^+=0$, a ``physical'' gauge for a system moving very fast 
along the ``$+$'' direction. In this gauge we evaluate the gluon scattering 
amplitude:
\beqar
M_{el}^g &=& \int \frac{d^4 q}{(2\pi)^4} \; 
[ (-g_s f_{acb})(g_{\alpha\mu} (p-q)_\beta + g_{\mu \beta}(2q-p)_\alpha 
+ g_{\beta\alpha}(-2p-q)_\mu ]  \frac{i}{ (p+q)^2 + i \epsilon }  
\nonumber \\[1ex] 
&& \times  \left[ -g^{\beta\beta^\prime}+ 
\frac{n^\beta (p+q)^{\beta^\prime} + n^{\beta^\prime} (p+q)^\beta}
{(p+q)\cdot n}  \right]  \, V_\mu^c(q) 
 M_{2, \beta^\prime}(p+q) M_{1,\alpha}(p)  
\nonumber \\[1ex]  
&\approx&  \int \frac{d q^-}{2\pi}\frac{d^2 {\bf q}_\perp}{(2\pi)^2} 
\left[\frac{2p^+ \, v(\vec{\bf q}_\perp)}{(p+q)^2+i\epsilon} \right]
 e^{iq^-(z-z_0)^+}  e^{-i{\bf q}_\perp \cdot 
({\bf z}_\perp-{\bf z}_{0\, \perp})}  \, T^c(A) \; T^c(T)  
\nonumber \\[1ex] 
&& \times \;  ( M_{2,\alpha} M_1^\alpha ) \;\;. \qquad 
\eeqar{g-el-amp} 
In the derivation of Eq.(\ref{g-el-amp}) we have used the approximations
$|{\bf q}_\perp| \ll p^+$ and $(p\pm q)^\mu M_\mu(p) \approx 0$.
The long range color interference has been neglected, allowing to factor 
out  $M_1^\alpha  M_{2,\alpha}$, and $-if_{acb}=[T^c(A)]_{ba}$ is the color 
matrix in the adjoint representation. We note that for small angle scattering 
of high-energy partons the amplitudes  $M_{el}^q, M_{el}^g$,  
Eqs.(\ref{q-el-amp},\ref{g-el-amp}), differ only by a color factor.

The remaining $q^-$ integral can be performed by closing the 
contour in the lower  half-plane ($z^+-z_0^+<0$) and picking the contribution 
at $q^- = {\bf q}^2_\perp/2p^+ - i \epsilon$, corresponding to the pole in 
the propagator of the scattered parton of momentum  $p+q$. The factorizable 
elastic scattering amplitude becomes:
\beq
M_{el}^{q,g} = -i \theta(z_0^+-z^+) \int \frac{d^2 {\bf q}_\perp}{(2\pi)^2} 
\; v({\bf q}_\perp) e^{i({\bf q}_\perp^2/2p^+)(z-z_0)^+} 
e^{-i{\bf q}_\perp \cdot ({\bf z}_\perp-{\bf z}_{0\, \perp})}
\, T^c(R=F,A)\; T^c(T) \;\; .  
\eeq{elamp}
Similarly, for final-state elastic scattering with the ``observed'' parton 
of momentum $p^\mu$, we get the same scattering amplitude except the pole of 
the $q^-$ integration is in the upper half-plane, 
$q^- = - {\bf q}^2_\perp/2p^+ + i \epsilon$, and the argument of the 
$\theta$-function in Eq.(\ref{elamp}) is reversed to $z^+-z_0^+$.

In computing the elastic scattering cross section in matter one takes 
the average over the distribution of scattering centers in the 
transverse plane of area $A_\perp$,  $1/A_\perp \int d^2 {\bf z}_\perp \, 
\langle \cdots \rangle$, which diagonalizes the squared amplitudes in the 
${\bf q}_\perp,{\bf q}_\perp^\prime $ 
variables~\cite{Gyulassy:2000fs,Gyulassy:2002yv}:      
\beq
\left \langle  e^{-i({\bf q}_\perp-{\bf q}_\perp^\prime)
\cdot({\bf z}_\perp - {\bf z}_{0\, \perp})} \right \rangle_{A_\perp} 
\simeq   \frac{T({\bf  z}_{0\, \perp})}{N} (2\pi)^2 
\delta^2({\bf q}_\perp-{\bf q}_\perp^\prime) \;\;
\eeq{expav}  
in the $\sqrt{A_\perp} \gg \mu^{-1}$ limit.  $T({\bf z}_{0\, \perp}) = 
\int dz \, \rho(z,{\bf z}_{0\, \perp})$  is the Glauber thickness function  
at impact parameter ${\bf z}_{0\, \perp}$ and  $ 1/A_\perp =  
T({\bf z}_{0\, \perp})/N $ with $N$ being  the number of scattering centers.  
The corresponding differential elastic scattering cross section per unit 
partonic scattering ($T({\bf z}_{0\, \perp})/N=1$) is found to be:
\beq
\frac{d\sigma_{el}(R,T)}{d^2{\bf q}_\perp}=
\frac{C_2(R)C_2(T)}{d_A}\frac{|v({\bf q}_\perp)|^2}{(2\pi)^2} 
= \frac{C_2(R)C_2(T)}{d_A} \frac{4 \alpha_s^2}{({\bf q}_\perp^2+\mu^2)^2}
\eeq{sigel}
for a fast parton in color representation $R$ and the non-Abelian
field radiated from a parton in color representation $T$.  
For example, the color factor $C_2(R)C_2(T)/d_A = 2/9, 1/2$, 
and 9/8 for the scattering of $qq$, $qg$, and $gg$, respectively.
A real nuclear medium can be a mixture of  soft quarks and gluons 
with a ``mean'' color factor $\langle C_2(R)C_2(T)/d_A\rangle$.
For $\mu=0$, Eq.(\ref{sigel}) is effectively the small angle,  
$t=-{\bf q}_\perp^2$ and $u \approx s$, limit of the exact 
$qq \rightarrow qq$, $qg \rightarrow qg$ and $gg \rightarrow gg$ 
elastic parton-parton scattering cross sections.

In deriving Eq.(\ref{sigel}) the ${\cal O}(1/p^+)$ correction,  
$\propto ({\bf q}_\perp^2/2p^+ -{\bf q'}_\perp^2/2p^+ )$, from the phase 
factors in Eq.(\ref{elamp}) vanishes due to Eq.(\ref{expav}), 
$\delta^2({\bf q}_\perp - {\bf q}_\perp^\prime)$. Thus, interference effects 
for an energetic jet are largely suppressed, but they become important 
for soft gluon radiation~\cite{Baier:1997kr,Gyulassy:2000fs}. In the case 
of multiple elastic scatterings the proper inclusion of unitarity corrections 
and their derivation has been discussed  in Ref.~\cite{Gyulassy:2002yv}. 
We present here the final result, noting that there is once again 
cancellation of the interference effects controlled by a phase 
$\propto ( {\bf q}_\perp + {\bf q}_\perp^\prime )^2/2p^+$ due to the forward 
weight, $\delta^2({\bf q}_\perp + {\bf q}_\perp^\prime)$, of the virtual 
scattering: 
\beq
\frac{d\sigma_{u.c.}(R,T)}{d^2{\bf q}_\perp} =
-  \left[ \int d^2 {\bf \xi}_\perp \frac{C_2(R)C_2(T)}{d_A} 
\frac{|v({\bf \xi}_\perp)|^2}{(2\pi)^2} \right] \delta^2({\bf q}_\perp)  
= -  \sigma_{el}\,  \delta^2({\bf q}_\perp) \;\; , 
\eeq{v-sigma}
i.e. the total scattering cross section in the forward ${\bf q}_\perp=0$ 
direction.

The full solution for the momentum distribution of a jet that has
traversed nuclear matter of longitudinal extent $L$ and opacity
$ \chi=L/\lambda  = \rho L \sigma_{el} = T({\bf z}_{0\;\perp}) 
\sigma_{el}$  can be obtained using the Gyulassy-Levai-Vitev reaction 
operator approach~\cite{Gyulassy:2002yv}: 
\beqar
\frac{d^3N^{f}(p^+,{\bf p}_\perp)}{d p^+ d^2 {\bf p}_\perp }
|_{ p^- = \frac{ {\bf p}_\perp^2}{2 p^+ } } 
& = &  \sum_{n=0}^{\infty} \frac{\chi^n}{n!} \int  \prod_{i=1}^n   
d^2 {\bf q}_{i\,\perp}  \left[  \frac{1}{\sigma_{el}} 
\frac{d\sigma_{el}(R,T)}{d^2{\bf q}_{i\,\perp}} \,
 \left(   e^{-{\bf q}_{i\, \perp} \cdot 
\stackrel{\rightarrow}{\nabla}_{{\bf p}_\perp} } 
 - 1  \right) \, \right]  \;   
\nonumber \\[1ex]
&&  \times  \frac{d^3N^{i}(p^+,{\bf p}_\perp)}{d p^+ d^2 {\bf p}_\perp } 
|_{ p^- = \frac{{\bf p}_\perp^2}{2 p^+ }}  \; 
\eeqar{ropit0}
for the leading power in ${\bf q}_\perp^2/P^2$.  Eq.(\ref{ropit0}) shows 
that in the asymptotic limit, ${\bf q}_\perp^2/P^2 \simeq 0$, elastic 
multiple parton scattering in nuclear matter introduces transverse momentum 
broadening to the jet without changing its ``+'' momentum. The average 
``$-$'' momentum component becomes $p^- = \langle {\bf p}_\perp^2 \rangle / 2p^+$
from the on-shell constraint  but  3D momentum is 
not conserved: $ | \vec{\bf p}^2| = \langle p_\parallel \rangle^2 +
\langle {\bf p}_\perp^2  \rangle  \approx 
(P^2 - \langle {\bf p}_\perp^2 \rangle /2 ) + \langle {\bf p}_\perp^2 \rangle 
\neq P^2 $.

The constant ``+'' momentum  component is a direct consequence of 
the leading power approximation, which results in the momentum constraint
$\delta(p'^+-p^+)=\delta(q^+)$ in Eq.(\ref{V-solution}). We here focus on 
the ${\bf q}_\perp^2/P^2$ power correction due to the medium recoil 
because it gives a leading contribution to the change of the fast parton's 
``+'' momentum. In a simple model of elastic jet scattering off a non-Abelian 
field radiated by a parton with momentum  $k^\nu=k^{\nu,-}=(\sqrt{2}P)n^\nu$ 
the correction from the medium  response replaces the momentum constraint 
at the leading power by $\delta(p'^+ - p^+(1- k'^+/p^+))$  for recoil 
momentum $k'^+$.  We find $k'^+= ({\bf q}^2_\perp/(p+k)^2) p^+
= \frac{1}{{2}} ({\bf q}^2_\perp/\sqrt{2}P)$ if the scattering puts the 
recoil parton on-shell ($k'^2=0$). By including this longitudinal momentum 
reduction Eq.(\ref{ropit0}) is modified to:
\beqar
\frac{d^3N^{f}(p^+,{\bf p}_\perp)}{d p^+ d^2 {\bf p}_\perp }
|_{ p^- =\frac{ {\bf p}_\perp^2}{2 p^+ } }  
& = &  \sum_{n=0}^{\infty} \frac{\chi^n}{n!} 
\int  \prod_{i=1}^n   d^2 {\bf q}_{i\,\perp} 
 \left[  \frac{1}{\sigma_{el}} 
\frac{d\sigma_{el}(R,T)}{d^2{\bf q}_{i\,\perp}} \,
 \left(   e^{-{\bf q}_{i\, \perp} \cdot 
\stackrel{\rightarrow}{\nabla}_{{\bf p}_\perp} } \, 
 e^{+\frac{1}{{2}} ({\bf q}^2_{i\,\perp} /\sqrt{2}P)\, 
\partial_{p^+} }  - 1  \right) \, \right]  \;   \nonumber \\[1ex]
&&  \times  
\frac{d^3N^{i}(p^+,{\bf p}_\perp)}{d p^+ d^2 {\bf p}_\perp }
|_{p^- =  \frac{ {\bf p}_\perp^2}{2 p^+ } }    \; .
\eeqar{ropit}
For any initial jet flux, Eq.(\ref{n0}), the opacity series 
in Eq.(\ref{ropit}) is most easily resummed in the impact parameter 
space $(b^-,{\bf b}_\perp)$ conjugate to $( p^+,{\bf p}_\perp )$. We 
substitute in Eq.(\ref{ropit}) the illustrative example of an initial 
parton momentum distribution, Eq.(\ref{fbeam}), which gives
\beqar
\left[ \frac{ d^3 \widetilde{{N}^{f}} }
{d p^+ d^2 {\bf p}_\perp } \right] (b^-,{\bf b}_\perp)
&=&  \frac{e^{-\chi} }{(2\pi)^2} \frac{e^{i \, \sqrt{2}P \, b^-}}{2 \pi}  
\; \exp \left[ \chi \int d^2 {\bf q}_\perp \, \frac{1}{\sigma_{el}} 
\frac{d\sigma_{el}}{d^2{\bf q}_{\perp}}  
e^{-i{\bf q}_{\perp}\cdot {\bf b}_{\perp} } 
e^{-i\frac{1}{{2}}({\bf q}^2_\perp/\sqrt{2}P)\, b^- }  \right] \;\;.  
\nonumber \\ && 
\eeqar{bspace}
While it is difficult to find a closed form for the integral in
Eq.(\ref{bspace}) even for simple forms of the differential 
elastic scattering cross section, the first correction for large
longitudinal momentum can be evaluated by expanding  
$e^{-i\frac{1}{{2}} ({\bf q}_{\perp}^2/{\sqrt{2} P}) b^-} = 
1 - i \frac{1}{{2}} ({\bf q}_{\perp}^2/{\sqrt{2} P}) b^- + \cdots $. 
The azimuthal integral in Eq.(\ref{bspace}) yields $J_0(b_\perp q_\perp)$ 
and the  first two terms in the expansion of the exponent integrated 
over the Yukawa potential give:
\beqar
\int_0^\infty dq_\perp \, q_\perp \frac{2\mu^2}{(\mu^2+q_\perp^2)^2} 
  J_0(b_\perp q_\perp) &=& b_\perp \mu \, K_1(b_\perp \mu) \;\;,
\nonumber \\[1ex]
-i \frac{1}{{2}}\frac{b^-}{\sqrt{2}P} 
\int_0^\infty dq_\perp \, 
q_\perp^3 \frac{2\mu^2}{(\mu^2+q_\perp^2)^2} 
  J_0(b_\perp q_\perp) 
&=&  \left( -i \frac{1}{{2}}\frac{b^-}{\sqrt{2}P} \right) 
 \mu^2( 2 K_0 (b_\perp \mu)
  - b_\perp \mu \, K_1(b_\perp \mu) ) \;\;. \nonumber \\[.5ex] && 
\eeqar{contribs}   
The key to evaluating the average ${\bf p}_\perp$-broadening and 
the shift in the ``+'' momentum 
of the partons is the small $b_\perp \mu$ expansion in 
Eq.(\ref{contribs}): $ b_\perp \mu \, K_1(b_\perp \mu) = 
1- (b_\perp^2 \mu^2/2)[  \ln (2e^{-\gamma_E}/(b_\perp \mu) ) + 1/2 ] 
+ {\cal O}(b_\perp^4 \mu^4)$,  $K_0 (b_\perp \mu) = 
\ln ( 2e^{-\gamma_E}/(b_\perp \mu )) + {\cal O}(b_\perp^2 \mu^2) $.
Keeping terms $ \propto \chi \mu^2 \xi$, where $\xi = 
\ln ( 2e^{-\gamma_E}/(b_\perp \mu )) \gton {\cal O}(1)$ we find:
\beq
\left[ \frac{d^3\widetilde{{N}^{f}}}{d p^+ d^2 {\bf p}_\perp } \right]
(b^-,{\bf b}_\perp)
=  \frac{1}{(2\pi)^3}\, e^{- \frac{1}{2}\chi \mu^2 \xi {\bf b}^2_\perp } 
\, e^{+i \left(\sqrt{2} P- \frac{1}{2} \frac{2 \chi \mu^2 \xi}
{ \sqrt{2} P}  \right) \, b^- }\;\;.
\eeq{bspace-approx} 
Because of the power behavior of $b_\perp^2$, we treat 
$\xi$ as approximately constant when Fourier transforming Eq.(15)
back to momentum space:
\beq
\frac{d^3N^{f}(p^+,{\bf p}_\perp)}{d p^+ d^2 {\bf p}_\perp }
|_{p^- =  \frac{ {\bf p}_\perp^2}{2 p^+ } } 
= \frac{1}{2\pi} \frac{e^{-\frac{{\bf p}^2_\perp}{2\,\chi \,\mu^2 \xi}}}
 {\chi\, \mu^2\, \xi } \,  \delta\left[p^+ - 
\left(\sqrt{2}P-\frac{1}{{2}} \frac{2 \chi \mu^2 \xi}{\sqrt{2} P}  
 \right) \right]  \;\;.
\eeq{gauss}  
The broadening of the parton beam, approximated in Eq.(\ref{gauss}) 
by a Gaussian, induces a negative light-cone component via the poles 
of the projectile propagators (see e.g. Eqs.(\ref{q-el-amp},\ref{g-el-amp})) 
that ensure on-shellness at all intermediate stages.  
The power corrections from the target recoil  lead to a reduction  
of the large ``+'' momentum component: 
$p^+ \rightarrow   p^+ -  \langle {\bf p}_\perp^2 \rangle/ 2 p^+ $, where 
$\langle {\bf p}_\perp^2 \rangle = \int d^2 {\bf p}_\perp \; 
{\bf p}_\perp^2 (d^2N^f/d^2{\bf p}_\perp) \, / \int d^2 {\bf p}_\perp \; 
(d^2N^f/d^2{\bf p}_\perp)  = 2\chi \mu^2 \xi$ is evaluated
from Eq.(\ref{gauss}). We note that the recoil power corrections 
to $p^-$ are ${\cal O}  
( \langle {\bf p}_\perp^2 \rangle^2 /(p^+)^2 ) $ and hereby neglected. 
To the order to which we
computed  the final state distribution  $\langle  {\bf p}^2_\perp \rangle 
= 2 \chi \mu^2 \xi$, $ \langle p^+ \rangle = \sqrt{2} P - \frac{1}{{2}} 
\frac{ 2 \chi \mu^2\xi}{\sqrt{2}P}$ and $ \langle p^- \rangle  
= \frac{1}{{2}} \frac{ 2 \chi \mu^2\xi}{\sqrt{2}P}$. 
We emphasize that proper inclusion of the target 
response  ensures  $p^0 = P$,  $ | \vec{\bf p}^2| =  \langle 
p_\parallel \rangle ^2 + \langle {\bf p}_\perp^2  \rangle  =P^2$ 
energy  and  3D momentum conservation in our final result
Eq.(\ref{gauss}).

One important consequence of the formalism that we have presented
is the longitudinal momentum  backward shift:
\beq
- \frac{dp_\parallel}{dz}  \approx - 
\frac{\Delta p_\parallel}{L} 
= \frac{  \mu^2 \, (2\xi) }{\lambda_{q,g}} \, \frac{1}{2 p_\parallel}  
= \left(\frac{\mu^2}{\lambda_{q,g}}\right)_{eff} 
\frac{1}{2 p_\parallel}  \;\; , 
\eeq{parshift}
that couples to the transverse momentum broadening, and may mimic 
small elastic  energy  loss if the full structure  of  
${d^3{N}^{f}}/{d p^+ d^2 {\bf p}_\perp }$  is not observed.    
We note that the results on  elastic transverse momentum diffusion 
presented in this Section are rather general, for example, the same 
expression is obtained for electron scattering via photon exchange or 
the scattering of a fast nucleon in nuclear matter via pion exchange.

\section{Application to the broadening of the back-to-back di-hadron 
correlation function}

As an application of the multiple initial and  final state elastic 
scattering formalism elaborated here we consider the nuclear induced  
broadening of the  back-to-back jet correlations associated with
hard QCD  $ab \rightarrow cd$ partonic subprocesses.  We  will limit 
the  discussion  to the Gaussian 2D random walk approximation, 
Eq.(\ref{gauss}), to make use of its additive dispersion property. 
Beyond this approximation the power law corrections to the tails of 
the distribution can be numerically evaluated as 
in~\cite{Gyulassy:2002yv}.

For $p+p$ collisions discussion of the jet acoplanarity resulting 
from vacuum radiation is given in~\cite{Feynman:yr}.  Experimental 
test of the theoretical estimates is performed through measurements 
of the near-side and away-side  di-hadron correlations in a plane 
perpendicular to the collision axis~\cite{Adler:2002tq,Angelis:1980bs}. 
In the absence of soft gluon bremsstrahlung and deviations from 
the double collinear pQCD approximation the back-to-back leading 
hadrons are expected to be coplanar. Two major effects contribute 
to the  observed acoplanarity in $p+p$: the non-perturbative 
fragmentation process, where final state hadrons pick up momentum 
${\bf j}_T$ perpendicular to the thrust axis, and the ${\bf k}_T$ 
dependence in the unintegrated parton distributions 
$\phi_{\alpha/h}(x,{\bf k}_T)$~\cite{Collins:1981uw}. 
Phenomenologically extracted  $\langle {\bf k}^2_T \rangle_{pp}$  
vacuum broadening can be  implemented  in the pQCD hadron 
production formalism  as  in~\cite{Feynman:yr,Owens:1986mp} via
is a normalized 2-dimensional Gaussian. The ${\bf j}_T$  dependence 
of $D_{h/\alpha}(z,{\bf j}_T)$ has not been discussed in the context 
of single inclusive particle production.

Measurements of intra-jet correlations find an approximately Gaussian 
jet cone shape. If one defines $\langle |{\bf j}_{T\,y}| \rangle$ 
to be the average particle transverse momentum  relative to  the hard 
scattered parent parton in the plane normal to the collision axis,  
it  can be  related  to the width $\sigma_{Near}$ of the near-side 
$(\Delta \phi < \pi/2)$ di-hadron correlation function  
$C(\Delta \phi) = N^{h_1,h_2}(\Delta \phi) / N_{tot}^{h_1,h_2}$ 
as follows:  $\langle | {\bf j}_{T\,y} | \rangle = 
\langle | {\bf p}_T | \rangle  \sin ( \sigma_{Near} / \sqrt{\pi}) $. 
Results from the CCOR collaboration (consistent with preliminary 
PHENIX measurements)~\cite{Angelis:1980bs}  give a value for 
$\langle | {\bf j}_{T\,y} | \rangle  \simeq 400$~MeV that is roughly 
independent of $\sqrt{s}$ and the $p_T$ of the trigger  particle.
This allows for the approximate separation of short and  long distance 
dynamics in QCD in accord with the factorization theorem~\cite{Collins:gx}
for inclusive hadron production with momentum exchange much larger than 
$\langle | {\bf j}_{T\,y} | \rangle$ and 
$\langle | {\bf k}_{T\,y} | \rangle$:
\beq
E \frac{d \sigma^h}{d^3 p}  \propto 
\sum_{ab,c} \phi(x_a,{\bf k}_{T\,b}) \otimes 
\phi(x_b,{\bf k}_{T\,b}) \otimes  \frac{d \sigma^{ab \rightarrow cd} }{dt}
\otimes D_{c/h}(z_c, {\bf j}_T)\;\;,
\eeq{facto}
where ``$\otimes$'' denotes standard integral convolution over the parton
momentum fractions. Moreover, a lack of statistically significant 
broadening of the {\em near-side} di-hadron correlations in $Au+Au$ 
can be interpreted as a signal for fragmentation sufficiently outside of 
the interaction region where the hot and dense quark-gluon 
plasma~\cite{CP} is expected to be formed. The reported consistency in 
the shape of the near-side peak of $C(\Delta \phi)$ in $p+p$, $d+Au$, 
and $Au+Au$ for $p_T \gton 2 $~GeV~\cite{Adler:2002tq,Angelis:1980bs} 
puts strong constraints on non-fragmentation and short formation time 
hadronic scattering models~\cite{Gallmeister:2002us} that, similarly 
to hydrodynamics~\cite{Huovinen:2003fa}, are free of jet-like 
correlations.

The out of trigger plane momentum component of the far-side 
correlated hadron is given by~\cite{Feynman:yr}: 
\beq
\langle |{\bf p}_{T\,out}| \rangle^2 
= \langle  | {\bf j}_{T\,y} |  \rangle^2  
+ x_E^2 \, ( \langle | {\bf j}_{T\,y} | \rangle^2 
+ 2 \,\langle | {\bf k}_{T\,y} | \rangle^2 ) \; .
\eeq{eq:ptout}
In Eq.(\ref{eq:ptout}) $x_E = -{\bf p}^h_{T} \cdot {\bf p}_{T\, trig}  
/ |{\bf p}_{T\, trig}|^2  \approx - \cos ( |\Delta \phi |  )$, 
where $ \Delta \phi$ is the angle between the 
approximately  back-to-back hadrons 
$h_1,h_2 $ and  $ \langle | \Delta \phi | \rangle = \sqrt{2/\pi} 
\,\sigma_{Far} $. 
Expressing also $ \langle |{\bf p}_{T\,out}| \rangle 
=  |{\bf p}_T^h |  \sin (|\Delta \phi |)$ and eliminating it in 
Eq.(\ref{eq:ptout}) for $\langle | \Delta \phi | \rangle$, 
$|{\bf p}_T^h| \approx | {\bf p}_{T\,trig} | 
= \langle |{\bf p}_{T}| \rangle$  the following approximate relation  
is found:
\begin{equation} 
\langle | {\bf k}_{T\,y} | \rangle = \langle |{\bf p}_{T}| 
\rangle \cos \left( \frac{\sigma_{Near}}{\sqrt{\pi}} \right)
\sqrt{ \frac{1}{2}\tan^2 \left(  \sqrt{\frac{2}{\pi}}\, \sigma_{Far} \right) 
- \tan^2 \left( \frac{\sigma_{Near}}{\sqrt{\pi}}   \right) } \;\; .
\label{eq:relat}
\end{equation}

In the presence of nuclear matter initial state (IS) and final state (FS) 
transverse momentum diffusion add a large $\langle \Delta {\bf k}_T^2 
\rangle  \propto  A^{1/3}$ term to the vacuum parton
broadening~\cite{Luo:ui,Gyulassy:2002yv}, as confirmed by Fermilab 
experiments on the nuclear $A$-dependence of di-jet 
acoplanarity~\cite{CORCO}. For $d+Au$ reactions at RHIC in the IS before 
the hard collision only the  partons from the incoming  deuteron scatter 
multiply on the nucleus: $\langle \Delta {\bf k}_T^2 \rangle_{IS} = 
(\mu^2/\lambda)_{eff} \langle L \rangle_{IS} $.  The gluon scattering 
dominated transport coefficient $\mu^2/\lambda \simeq 
0.14$~GeV$^2$/fm is constrained~\cite{Vitev:2002pf} from existing low 
energy $p+A$ data~\cite{Cronin:zm} and consistent with 
$\mu^2/\lambda_q = 0.047 \pm 0.035$~GeV$^2$/fm  found  
in~\cite{Arleo:2002ph} when one takes into account that
$\lambda_q/\lambda_g = C_A/C_R = 2.25$. After the hard collision 
vertex in the FS both outgoing  jets scatter in the  medium to 
acquire $\langle \Delta {\bf k}_T^2 \rangle_{FS}$.
Projection on the plane of measurement with which ${\bf k}_{T\; FS}$  
forms an angle $\alpha_{FS}$,  $\langle \cos^2 \alpha_{FS} \rangle = 1/2$,  
reduces the  FS  effect by a factor of two. Comparison of nuclear 
broadening in Drell-Yan and di-jet production indicate that the strength 
of FS scattering may be  bigger in the IS~\cite{CORCO}. 
This can be modeled  by $K_{FS} = (\frac{\mu^2}{\lambda})_{eff \; FS} 
/ (\frac{\mu^2}{\lambda})_{eff \;IS} $. Lacking precise experimental
data, we naturally choose $K_{FS} = 1$ to be the default value 
of this parameter. The total vacuum+nuclear induced broadening in the  
plane perpendicular to the collision axis in $d+Au$ is given by:
\beq
 \langle  {\bf k}_T^2 \rangle = \langle  {\bf k}_T^2 \rangle_{vac} +   
1_{jet} \, \left( \frac{\mu^2}{\lambda} \right)_{eff} 
\langle L \rangle_{IS} 
+  2_{jets}\,  \left(\frac{1}{2} \right)_{projection}   \,  K_{FS} \,
\left( \frac{\mu^2 }{\lambda} \right)_{eff} \langle L \rangle_{FS}\;\;.
\eeq{eq:netbr}
To  relate the effective opacity (or mean number of scatterings) 
$\chi = \bar{n} =  \langle L \rangle / \lambda_{eff}$  for  the  
IS and the FS  it is useful to look in the rest frame of the nucleus 
where the back-to-back midrapidity hard scattered partons  move 
in the forward direction at an angle $\theta \approx \gamma^{-1}$ relative 
to the incident beam axis, $\gamma = E_N/m_N = \sqrt{s}/(2m_N)$ 
being the Lorentz boost factor. For RHIC $\gamma=106$ and uniform 
distribution of the hard scatter in the nucleus $\chi_{IS} = \chi_{FS}$.

\begin{figure}[t!]
\begin{center} 
\vspace*{-0.1in}
\psfig{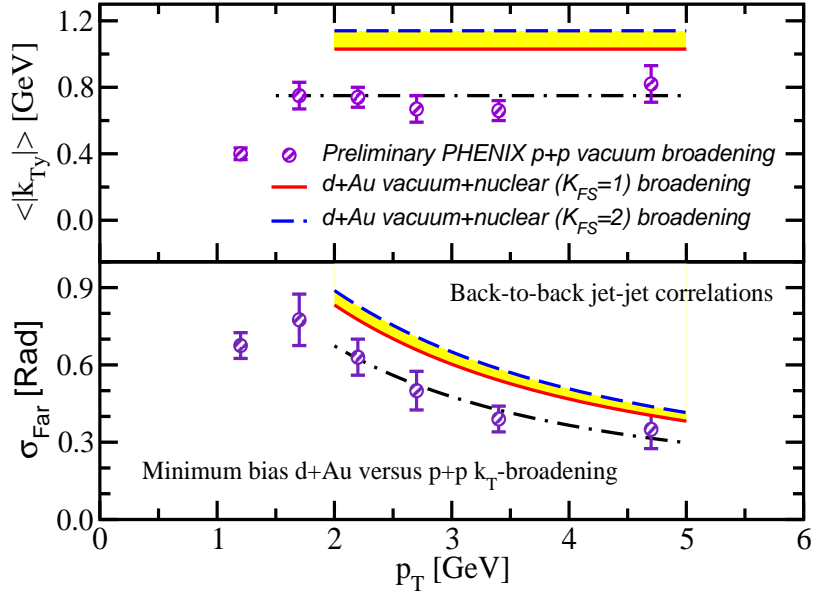}
\vspace*{-0.1in}
\caption{Predicted broadening of the  back-to-back jet correlations 
in minimum bias $d+Au$ reactions relative to the vacuum case 
is illustrated for $\langle | {\bf k}_{Ty} |  \rangle$ (top panel)  
and  $\sigma_{Far}$ (bottom panel). The strength of FS interactions 
is taken to be $K_{FS}=1,2$ times bigger than in the IS and 
represented by a band. $K_{FS} = 2$ is a good estimate of 
the different amount of ${\bf k}_T$-diffusion in central versus minimum 
bias events. Preliminary $p+p$ data is from PHENIX and the 
dot-dashed line estimates the vacuum broadening.}
\label{broad:fig2}
\end{center} 
\end{figure}

The estimate for $\langle | {\bf k}_{T\,y} | \rangle_{vac} = 0.75$~GeV  
per parton
is taken from~\cite{Angelis:1980bs} (shown by dot-dashed line in 
Fig.~\ref{broad:fig2}) and the di-hadron correlation function is 
approximated by near-side and far-side Gaussians. Such simplification 
does not take into account the large angle multiple scattering and 
fragmentation and the growth of the background in the 
intermediate $\Delta \phi \sim \pi/2$ region.  The combined 
vacuum+nuclear matter induced IS  and FS transverse momentum diffusion 
for minimum bias $d+Au$ reactions is evaluated from Eq.(\ref{eq:netbr}).
Noting that for two Cartesian components  $\langle |{\bf k}_{T\,y}| 
\rangle  = \sqrt{\langle {\bf k}_{T}^2 \rangle/\pi}$, 
in the top panel of Fig.~\ref{broad:fig2} $\sim 30 \%$ enhancement of  
$\langle | {\bf k}_{T\,y} | \rangle $ that measures the increased 
acoplanarity due to $k_T$-broadening is predicted. 
The lower panel illustrates the increased width, $\sigma_{Far}$, of 
the away-side correlation function  $C(\Delta \phi )_{dAu}$ obtained
as a solution to Eq.(\ref{eq:relat}).

\begin{figure}[t!]
\begin{center} 
\vspace*{-0.1in}
\psfig{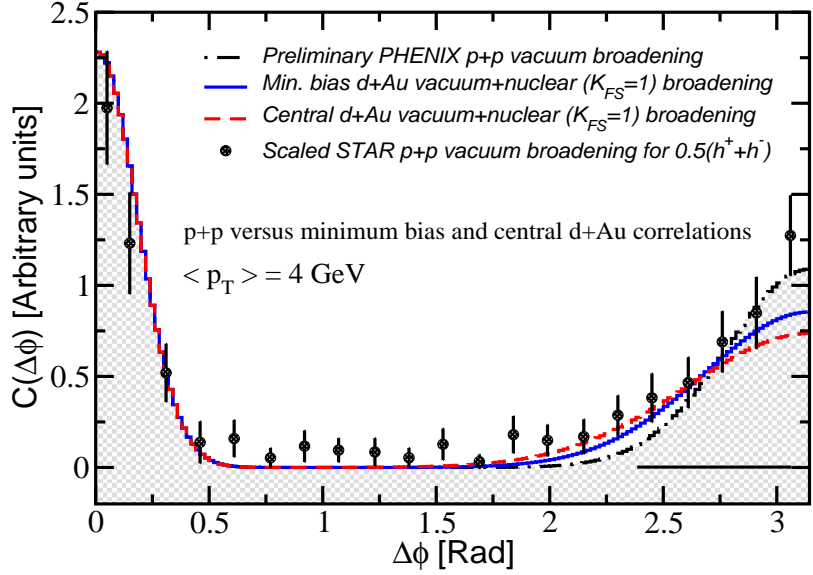}
\vspace*{-0.1in}
\caption{Elastic broadening of the back-to-back di-hadron correlation 
function in minimum bias and central $d+Au$ reactions relative to the 
vacuum $p+p$ result (here represented by a shaded area to guide the eye)
for $\langle | {\bf p}_T |  \rangle = 4$~GeV. $C(\Delta \phi)$ is 
modeled via near-side and far-side Gaussians with 
away-side width evaluated from Eqs.(\ref{eq:relat},\ref{eq:netbr}) 
and $K_{FS}=1$. Scaled STAR correlation data~\cite{Adler:2002tq} 
for  $\half (h^+ + h^-)$ in $p+p$ is also shown.} 
\label{broad:fig3}
\end{center} 
\end{figure}

Figure~\ref{broad:fig3} shows the change in the shape of the far-side 
di-hadron correlations from transverse momentum broadening and the  
coupled $~20\%$ reduction of their maximum strength 
(at $\Delta \phi = \pi $). In central $d+Au$ reactions the effect of 
nuclear matter induced  acoplanarity is slightly larger. We emphasize 
that even a significant increase in the opacity of the medium  will 
lead to only a subtle change in $\sigma_{Far}$ since its observable 
growth is no faster than $\sqrt{\chi}$. Triggering on very 
high-$p_T$ hadrons will bias the measurement toward quark jets and 
may lead to a reduction of the estimated broadening.

For  $Au+Au$ collisions one has to account for the 
IS broadening of both partons and replace the FS diffusion in cold 
nuclear matter with quark-gluon plasma induced broadening, which 
for the Bjorken expansion scenario~\cite{Huovinen:2003fa} 
at midrapidity has the form: 
\beq
\langle  \Delta {\bf k}_T^2 \rangle_{FS} = C_2(R)  
\frac{3 \pi \alpha_s^2}{2} \frac{1}{A_\perp} \frac{dN^{g}}{dy}
\, 2\xi \,    
\ln \frac{  \langle L_T  \rangle }{\tau_0}\;,  
\qquad C_2(R) = 4/3\; (3) \;\; {\rm for\; quarks\; (gluons)} \;\;.
\eeq{bjbroad} 
In Eq.(\ref{bjbroad}) ${dN^{g}}/{dy}$ is the initial gluon rapidity
density, $\langle L_T \rangle$ is the
transverse size of the medium and ${\tau_0}$ is the initial 
equilibration time~\cite{Huovinen:2003fa}.

\section{Conclusions}

We have extended the GLV reaction operator formalism to multiple 
elastic scatterings~\cite{Gyulassy:2002yv} to take into account the 
energy-momentum conservation for the incident parton flux and the 
recoil of the target partons along the projectile path. The resulting 
coupling between the forward and backward light-cone and transverse 
momentum components of the distribution of jets that have traversed 
dense nuclear matter is non-negligible for small jet energies. We have 
introduced a boost invariant formulation of the problem of multiple 
parton interactions that generalizes the Gyulassy-Wang model of 
dense nuclear matter~\cite{Gyulassy:1993hr}. Relaxing the assumptions 
of static or very massive scattering centers in the target 
simplifies the calculation, the transition between different 
reference frames and also allows to analytically include 
corrections, accurate to ${\cal O}(1/p^+)$ in Eq.(\ref{gauss}), 
to the leading power approximation.

In a practical application of the transverse  momentum  diffusion  
results we evaluate the nuclear-induced acoplanarity of the 
hard-scattered parton pair, Eq.(\ref{eq:netbr}), and  relate it 
to the expected broadening of the back-to-back di-hadron correlation 
function in $d+Au$ reactions at RHIC. We find that scattering in cold 
nuclear matter, which results in a predicted small 20-30\%  Cronin 
enhancement~\cite{Vitev:2002pf}, also leads to $\sim 25-30\%$ 
increase of the far-side width $\sigma_{Far}$ of 
$C(\Delta \phi)_{dAu}$ in minimum bias $d+Au$ reactions for 
$\langle | {\bf p}_T | \rangle = 3-4$~GeV. The centrality 
dependence of $\sigma_{Far}$ is shown to be weak with $\sim 35-40\%$ 
growth of the far-side width in central $d+Au$ relative to $p+p$.  
In summary, di-hadron correlations for scattering in cold nuclear matter 
are predicted in Fig.~\ref{broad:fig3} to be qualitatively similar 
to the $p+p$ case. 

{\em Note added in proof:}  During the completion of this manuscript 
data on the nuclear modifications of hadron production and back-to-back 
di-hadron correlations in $d+Au$ at $\sqrt{s}_{NN}= 200$~GeV became 
available~\cite{Adler:2003ii}. The moderate-$p_T$ 
enhancement observed  in minimum bias $d+Au$ versus the jet quenching 
established in central and semi-central 
$Au+Au$~\cite{Adler:2003qi,Adler:2003ii} are in 
quantitative agreement with the predictions in~\cite{Vitev:2002pf}. 
$C(\Delta \phi)_{dAu}$  and $C(\Delta \phi)_{pp}$ have the same 
qualitative behavior in both the near-side and far-side regions. 
Further analysis of the experimental data is needed to clarify the 
centrality and rapidity dependence~\cite{Vitev:2002pf,Gelis:2002nn} of 
the Cronin effect at RHIC and to quantify the broadening of the 
far-side width, $\sigma_{Far}$, discussed here.                 

\vspace*{-0.1in}   
\begin{acknowledgments}
Many illuminating discussions  with  M.~Gyulassy and  J.P.~Vary  are  
gratefully  acknowledged. We thank P. Constantin, J. Lajoie, C. Ogilvie 
and J. Rak for highlighting the importance of $d+Au$ acoplanarity 
measurements and  D. Hardtke, P. Jacobs and M. Miller for clarifying 
comments on STAR  di-hadron correlations data. We acknowledge the 
hospitality of the nuclear theory groups at Columbia University (I.V.)  
and the Institute for Nuclear Theory (J.Q. and I.V.) where parts of this 
manuscript were completed. This work is supported in part by the United 
States Department of Energy under Grants  No. DE-FG02-87ER40371 and  
No. DE-FG02-93ER40764. 
\end{acknowledgments}



\end{document}